\documentstyle[12pt,aps]{revtex}
\draft

\begin{document}

\title{Many-body $q$-exponential distribution prescribed by factorization
of joint probability}

\author{Qiuping A. Wang}
\address{Institut Sup\'erieur des Mat\'eriaux du Mans,\\ 44, Avenue F.A.
Bartholdi, 72000 Le Mans, France}

\date{\today}

\maketitle

\begin{abstract}
The factorization problem of $q$-exponential distribution within
nonextensive statistical mechanics is discussed on the basis of
Abe's general pseudoadditivity for equilibrium systems. It is
argued that the factorization of compound probability into product
of the probabilities of subsystems is nothing but the consequence
of existence of thermodynamic equilibrium in the interacting
systems having Tsallis entropy. So the factorization does not
needs independent noninteracting systems and should be respected
in all exact calculations concerning interacting nonextensive
subsystems. This consideration makes it legitimate to use
$q$-exponential distribution either for composite system or for
single body in many-body systems. Some known results of ideal
gases obtained with additive energy are reviewed.
\end{abstract}



\section{Introduction}

In thermostatistics, the factorization of the compound probability
into product of single body probabilities
\begin{eqnarray}                                        \label{m1}
\rho=\prod_{n=1}^N\rho_n,
\end{eqnarray}
where $\rho$ is density operator and $N$ is the number of bodies
in the system, is crucial for applications of statistical
mechanics to many-body systems and for the statistical
interpretation of thermodynamics. Boltzmann-Gibbs statistics
($BGS$), in considering only short range interactions, makes this
factorization a natural result of its exponential distribution
with additive energy. In nonextensive statistical mechanics
($NSM$)\cite{Tsal88}, intended originally to describe complex
systems with long range interactions or fractal structure of
space-time, this factorization is not as evident as in $BGS$. From
the beginning of this nonextensive theory, in order to show the
nonextensive character of $NSM$, one has supposed Eq.(\ref{m1})
for a system containing {\it statistically independent subsystems}
and obtained for the total entropy $S_q$\cite{Tsal88} :
\begin{eqnarray}                                        \label{m2}
\ln[1+(1-q) S_q]=\sum_{n=1}^N \ln[1+(1-q) S_q(n)]
\end{eqnarray}
where $q$ is the parameter of Tsallis entropy
$S_q=-\texttt{Tr}\frac{\rho-\rho^q}{1-q}$ (Boltzmann constant $k_B=1$) [for
$N=2$, $S_q=S_q(1)+S_q(2)+(1-q) S_q(1)S_q(2)$ as one often finds in the
literature]. As indicated by Tsallis\cite{Tsal88}, Eq.(\ref{m2}) expresses in
fact the additivity of R\'enyi entropy
$S^R=-\frac{\texttt{Tr}\rho^q}{1-q}$\cite{Reny66} if Eq.(\ref{m1}) applies. Due
to this supposed ``independence" of subsystems for $NSM$, it has been believed
by many that exact calculations \cite{Tsal94,Prat94,Curi96,Lenz98,Marti,Lenz01}
should use the additive hamiltonian $H_0$ given by
\begin{equation}                                        \label{1}
H_0=\sum_{n=1}^N H_n,
\end{equation}
where $H_n$ is the hamiltonian of $n^{th}$ subsystem. However, this additive
hamiltonian is compatible with neither Eq.(\ref{m1}) nor Eq.(\ref{m2}) since
these relations applied to $q$-exponential distribution ($qed$)
$\rho\propto[1-(1-q)\beta H]^{\frac{1}{1-q}}$, as given by maximization of
Tsallis entropy under some constraints\cite{Tsal88,Marti,Wang00},
imply\cite{Tsal88,Wang01,Beck00}:
\begin{eqnarray}                                        \label{2}
H &=& \sum_{n=1}^N H_n + \sum_{k=2}^N [(q-1)\beta]^{k-1}
    \sum_{n_1<n_2<...<n_k}^N\prod_{j=1}^k H_{n_j} \\ \nonumber
&=& H_0+H_c,
\end{eqnarray}
where $\beta$ is the inverse temperature. So in order to keep simultaneously
Eq.(\ref{1}) and Eq.(\ref{m1}) and to apply $NSM$ to many-body systems using
one-body $qed$, a so called {\it factorization approximation} is
proposed\cite{Buyu93} with the assumption that the second term on the right
hand side of Eq.(\ref{2}) may be neglected. This approximation has been,
explicitly or not, employed in the applications of $NSM$ to the cases like,
among others, quantum particle systems\cite{Buyu93,Kani96,Bedi01}, turbulent
flows\cite{Beck00,Beck01a}, the polytropic model of galaxies, solar neutrinos,
peculiar velocity of galaxy clusters, electron plasma (for updated comments on
some of these works, we refer to reference \cite{Tsal01} and the references
there-in). In these calculations, it is admitted that one-body $qed$ applies.
Although these applications clearly shows the usefulness and necessity of
one-body $qed$, the approximation\cite{Buyu93} neglecting the correlation
energy with sometimes weak interacting dilute particles\cite{Beck01} is, on the
contrary, not a reassuring basis. Recently, it has been argued\cite{Wang01}
that the correlation energy ($H_c$) given by the second term of Eq.(\ref{2}) is
in general not negligible. A numerical result for $N$-oscillator
system\cite{Lenz01} shows that the partition function given by using
Eq.(\ref{1}) is completely different from that given by using Eq.(\ref{2}) when
$N$ is large. We are facing with a difficult question : which one of
Eq.(\ref{m1}) and Eq.(\ref{1}) should be related to independence of subsystems?
In a recent work, Beck proposed a nontrivial idea to define an ``independence"
according to Eq.(\ref{2})\cite{Beck01} so we can write Eq.(\ref{m1}) and
one-particle $qed$ without any approximation. This means that we address
correlated ``independent systems" satisfying Eq.(\ref{2}). This idea surely
needs to be justified.

In this letter, the probability factorization problem of $NSM$ is
discussed from a different consideration inspired by the general
pseudoadditivity of entropy\cite{Abe01} and energy\cite{Wang02}
required by the existence of thermodynamic equilibrium. It will be
argued that Eq.(\ref{m1}) needs not independence of noninteracting
systems and is to be considered as a basic assumption of $NSM$.
Eq.(\ref{1}) is only a kind of extensive approximation and should
be employed carefully. On this basis, some basic applications of
$NSM$ to classic and quantum gases are revisited.

\section{Thermal equilibrium and factorization of compound
probability}

Apart from one-particle $qed$, another issue tightly related to probability
factorization is the establishment of zeroth law and the definition of
temperature for $NSM$. This is obviously of central importance for the theory.
Eq.(\ref{1}) and noninteracting model due to Eq.(\ref{m1}) are so deeply
accepted that it was even believed that the zeroth law of thermodynamics was
absent within $NSM$\cite{Guer96}, since Eq.(\ref{m2}) does not hold with
additive hamiltonian, and without Eq.(\ref{m2}), we can not talk about the zero
law and temperature! Recently, a series of works have been published on this
issue\cite{Abe00} claiming the establishment of zeroth law and the definition
of a generalized temperature on the basis of additive hamiltonian Eq.(\ref{1})
and, I stress it, Eq.(\ref{m2}). If this zeroth law is established by
neglecting the correlation energy $H_c$, it is merely approximate. If it is
exact for noninteracting systems, then the paradox between Eq.(\ref{1}) and
Eq.(\ref{m1}) [thus Eq.(\ref{m2})] persists due to $qed$. In addition, other
questions arise : if the zeroth law holds only for, e.g. noninteracting
particle systems, why do we have to discard $BGS$? And what is the origin of
the nonextensivity?

This confused situation and the above questions are, in our
opinion, simply due to the fact that Eq.(\ref{m1}) is not clearly
founded for $NSM$. It certainly implies independence of
noninteracting systems for $BGS$. But does it mean the same thing
for $NSM$?

Very recently, Abe\cite{Abe01} proposed a general pseudoadditivity
for entropy required by the existence of thermal equilibrium in
composite nonextensive systems : $f(S)=f(S_1)+f(S_2)+\lambda
f(S_1)f(S_2)$ where $f$ is certain differentiable function
satisfying $f(0)=0$ and $\lambda$ a constant depending on the
nature of the system of interest. So for a system containing $N$
subsystems, the thermal equilibrium requires following pseudoadditivity
:
\begin{eqnarray}                                        \label{2b}
\ln[1+\lambda f(S)]=\sum_{n=1}^N\ln[1+\lambda f(S_n)].
\end{eqnarray}
On the other hand, Eq.(\ref{2b}) applied to Tsallis entropy means
$f(S)=S$ and $\lambda=1-q$\cite{Abe01}, which directly leads to
$\ln\texttt{Tr}\rho^q=\sum_{i=1}^N\ln\texttt{Tr}\rho_i^q$ or
Eq.(\ref{m1}) (i.e. $(p_ip_j)^q=p_{ij}^q$ means $p_ip_j=p_{ij}$
where $p_i$ is the probability of state $i$). So this probability
factorization must be regarded as a condition or a consequence of
thermodynamic equilibrium, instead of statistical independence. It
should be raised to the rank of {\it basic assumption for
equilibrium thermodynamics with Tsallis entropy} and must be
rigorously respected by all exact calculations. In this way, the
zeroth law becomes evident and a temperature can be
straightforwardly defined with maximum entropy and minimum
energy\cite{Wang00,Wang01,Wang02a}. According to above
discussions, all calculations based on Eq.(\ref{m1}) or using
one-body $qed$ are, as a matter of fact, exact applications of
$NSM$. And all calculations based on additive energy Eq.(\ref{1})
should now be considered as a kind of ``extensive approximation"
and should be employed with great care.

Energy has been proved\cite{Wang02} to satisfy the same kind of
pseudoadditivity at equilibrium as Eq.(\ref{2b}). If we choose
$f(H)=H$ and $\lambda=(q-1)\beta$, we get
\begin{eqnarray}                                        \label{2c}
\ln[1+(q-1)\beta H]=\sum_{n=1}^N\ln[1+(q-1)\beta H_n]
\end{eqnarray}
which is just Eq.(\ref{2}) satisfying Eq.(\ref{m1}).

Here I would like to mention that, Eq.(\ref{2c}) or Eq.(\ref{2}) contains
inverse temperature $\beta$. This surely has something to do with the fact that
this relation is subject to thermal equilibrium. The dependence of the
correlation energy $H_c$ on temperature is something of nontrivial and needs
explanation. A possible argument is that $H$ is only an effective Hamiltonian
so the effects of interactions or correlations represented by $q$ perhaps
depends on the state of the system and thus on temperature. If this is true, we
may expect $T$-dependent $q$ values.

In the following section, I will discuss some interesting
consequences of Eq.(\ref{2c}).

\section{Additive $q$-deformed energy}
Let us begin by introducing an additive ``energy" for $NSM$ for the
sake of convenience. We refer to
\begin{eqnarray}                                        \label{11b}
h=\frac{\ln(1+(q-1)\beta H)}{(q-1)\beta}
\end{eqnarray}
as $q$-deformed hamiltonian and recast Eq.(\ref{2c}) into
\begin{eqnarray}                                        \label{12}
h=\sum_{n=1}^Nh_n.
\end{eqnarray}
This means following transformations :
\begin{eqnarray}                                        \label{13}
H=\frac{e^{(q-1)\beta h}-1}{(q-1)\beta }, \;\;\;
\;\;\;H_n=\frac{e^{(q-1)\beta h_n}-1}{(q-1)\beta }
\end{eqnarray}
and
\begin{eqnarray}                                        \label{14}
\rho\propto [1+(q-1)\beta H]^{1/(1-q)}= e^{-\beta h}
=\prod_ne^{-\beta h_n}.
\end{eqnarray}
It is interesting to see that the $qed$ can be transformed into Boltzmann-Gibbs
exponential distribution with a deformed energy. The reader may wonder why we
refer to the additive hamiltonian $h$ or $h_n$, instead of the nonadditive one
$H$ or $H_n$, as deformed hamiltonian. It should be noticed that, when
addressing a system of $N$ particles, we have to write
$H_n=\frac{p_n^2}{2m}+V_n$ for single particle so that
$h_n=\frac{\ln[1+(q-1)\beta(\frac{p_n^2}{2m}+V_n)]}{(q-1)\beta}$. It is clear
that $H_n$, instead of $h_n$, is the physical energy. According to this
consideration, the total hamiltonian of a system of "ideal gas" ({\it free
particles in the sense that we do not write the correlation energy between the
particles in Hamiltonian and let it be represented by $q\neq 1$ in the deformed
energy}) should be written as
\begin{eqnarray}                                        \label{15}
H=\frac{e^{(q-1)\beta
h}-1}{(q-1)\beta}=\frac{e^{(q-1)\beta\sum_{n=1}^Nh_n}-1}{(q-1)\beta}
=\frac{e^{\sum_{n=1}^N\ln[1+(q-1)\beta\frac{p_n^2}{2m}]}-1}{(q-1)\beta}
=\sum_{n=1}^N\frac{p_n^2}{2m}+H_c
\end{eqnarray}
When $q=1$ ($H_c=0$), we recover $H=\sum_{n=1}^N\frac{p_n^2}{2m}$.

I would like to emphasize here that, as shown in Eq.(\ref{14}), the deformed
hamiltonian $h$ may be employed, if we want, to recover the mathematical
structure of $BGS$ in introducing a $q$-deformed information measure
$I_q=-\ln\rho$ and an entropy $\mathcal{S}=-\texttt{Tr}\rho\ln\rho$ where
$\rho=\frac{e^{-\beta h}}{Z}$ ($Z=\texttt{Tr}e^{-\beta h}$). We
straightforwardly obtain $\frac{\partial \mathcal{S}}{\partial u}=\beta=1/T$
and a deformed first law $du=dw+Td\mathcal{S}$, where $u=\texttt{Tr}\rho h$ is
the deformed internal energy and $dw$ the work on the system.

\section{Review of some results about ideal gases}
Now I will show some consequences of Eq.(\ref{15}) on ideal gases.

\subsection{Classical ideal gas}
For classical ideal gas, the total hamiltonian should be given by
Eq.(\ref{15}). So the total partition function $Z$ is given by [in
the formalism of complete distribution ($\texttt{Tr}\rho=1$) with
unnormalized average energy $U=\texttt{Tr}H\rho^q$] :
\begin{eqnarray}                                        \label{17}
Z &=&\texttt{Tr}[1-(1-q)\beta H]^\frac{1}{1-q} \\\nonumber
&=&\prod_{n=1}^N\texttt{Tr}[1-(1-q)\beta H_n]^\frac{1}{1-q}
\\\nonumber
&=&\{\texttt{Tr}[1-(1-q)\beta H_n]^\frac{1}{1-q}\}^N=z^N
\end{eqnarray}
where $z=\texttt{Tr}[1-(1-q) H_n]^\frac{1}{1-q}$ is the single
particle partition function. The calculation of $Z$ in reference
\cite{Tsal94,Prat94} can be adopted for single particle. For
$q>1$, we obtain :
\begin{eqnarray}                                        \label{18}
z &=& \frac{\frac{(2\pi
m)^{d/2}V}{h^d}\Gamma(\frac{1}{q-1}-\frac{d}{2})}
{\Gamma(\frac{1}{q-1})(q-1)^{d/2}}\beta^{-d/2}=\Delta \beta^{-d/2}
\end{eqnarray}
where $d$ is the dimension of volume $V$. The total internal
energy is
\begin{eqnarray}                                        \label{19}
U &=&-\frac{\partial}{\partial\beta}\frac{Z_q^{1-q}-1}{1-q}
\\\nonumber &=&-\frac{\partial}{\partial\beta}
\frac{z_q^{N(1-q)}-1}{1-q} \\\nonumber
&=&-\frac{\partial}{\partial\beta} \frac{z_q^{N(1-q)}-1}{1-q}
\\\nonumber &=& \frac{dN}{2\Delta^{N(q-1)}}(kT)^{dN(1-q)/2+1}.
\end{eqnarray}
The specific heat is so given by
\begin{eqnarray}                                        \label{20}
C_v =\frac{\partial U}{\partial T}
=\frac{kdN}{2\Delta^{N(q-1)}}(kT)^{dN(1-q)/2}.
\end{eqnarray}
We notice that the $T$-dependence of $C_v$ is the same as given by the {\it
extensive approximation} but the $q-$ and $N$-dependences are
changed\cite{Tsal94}. This point is also noticed in the case of
$N$-harmonic-oscillator system\cite{Lenz01}. Another interesting point is that,
in the present case, the variation interval of $q$ is enlarged to
$1<q<1+\frac{2}{d}$ from $1<q<1+\frac{2}{dN}$ given by the extensive
approximation\cite{Tsal94,Prat94}.

For $q<1$, similar discussions can be made with the extension of
Hilhorst formula given by Prato\cite{Prat94}.

\subsection{Quantum gas}
For nonextensive quantum gas, from the $qed$ for grand canonical
ensemble\cite{Buyu93,Curi96} with complete distribution and
normalized average ($U=\texttt{Tr}H\rho$), we can write :
\begin{eqnarray}                                        \label{21}
[1+(1-q)\beta(H-\mu N)]^{\frac{1}{q-1}}=
\prod_{n=1}^N[1+(1-q)\beta(H_n-\mu)]^{\frac{1}{q-1}}.
\end{eqnarray}
Let this equation be replaced by
\begin{eqnarray}                                        \label{22}
e^{-\beta h'}=\prod_{n=1}^Ne^{-\beta h'_n}
\end{eqnarray}
where $h'=\frac{\ln[1+(1-q)\beta(H-\mu N)]}{\beta(1-q)}$ and
$h'_n=\frac{\ln[1+(1-q)\beta(H_n-\mu)]}{\beta(1-q)}$. The total
partition function $Z$ is then given by
\begin{eqnarray}                                        \label{23}
Z &=&\texttt{Tr}[1-(q-1)\beta(H-\mu N)]^\frac{1}{q-1}
\\\nonumber &=&\texttt{Tr}e^{-\beta h'} \\\nonumber
&=&\texttt{Tr}e^{-\beta \sum_nh'_n} \\\nonumber
&=&\prod_k\sum_{n_k}e^{-n_k\beta e'_k}
\end{eqnarray}
where $e'_k$ is the eigenvalue of $h'_n$ and $n_k$ the occupation
number of the state $k$ of single particle. For boson and
fermion, we obtain, respectively,
\begin{eqnarray}                                        \label{25}
Z=\prod_k\sum_{n_k=0}^\infty e^{-n_k\beta e'_k}
=\prod_k\frac{1}{1-e^{-\beta e'_k}} \;\;\;and\;\;\;
Z=\prod_k\sum_{n_k=0}^1 e^{-n_k\beta e'_k} =\prod_k(1+e^{-\beta
e'_k}).
\end{eqnarray}
Then, it is straightforward to get :
\begin{eqnarray}                                        \label{26}
\bar{n}_l=\texttt{Tr}\rho n_l=-\frac{1}{\beta}\frac{\partial(\ln
Z)}{\partial e'_l} =\frac{1}{e^{\beta e'_l}\pm1}
=\frac{1}{[1+(1-q)\beta(e_l-\mu)]^{\frac{1}{1-q}}\pm1}
\end{eqnarray}
where $e_l$ is the eigenvalue of the one particle hamiltonian
$H_n$. ``+" and ``-" correspond to fermions and bosons,
respectively. These are just the standard quantum distributions
given by B\"uy\"ukkili\c{c} {\it et al} with the so called
factorization approximation\cite{Buyu93} and unnormalized average.

On the other hand, in the present formalism of $NSM$, unnormalized
average does not give simple quantum distribution similar to
Eq.(\ref{26}) for standard bosons and fermions. This is due to the
fact that $Z=\prod_k\sum_{n_k} e^{-n_k\beta e'_k}\neq
\prod_k[\sum_{n_k} e^{-qn_k\beta e'_k}]^{1/q}$. Here I only show
that, in this case, Eq.(\ref{26}) becomes
\begin{eqnarray}                                        \label{26a}
\bar{n}_l=\texttt{Tr}\rho^q n_l
=\frac{Q}{[1+(q-1)\beta(e_l-\mu)]^{\frac{q}{q-1}}\pm1}
\end{eqnarray}
where $Q=\texttt{Tr}\rho^q$ can be regarded as a parameter
depending on $q$. $Q>1$ $Q=1$ and $Q<1$ for $q<1$, $q=1$ and
$q>1$, respectively. These distributions seems interesting because
they allow intermediate occupation number between that of bosons
and fermions. In particular, at absolute zero, for ``fermionlike"
particle with ``+", $\bar{n}_l=Q$ when $e_l<\mu$ and $\bar{n}_l=0$
when $e_l>\mu$. This means that it is possible for several
``fermions" to occupy an one-particle quantum state if $Q>1$ or
$q<1$. Consequently, the Fermi surface $\epsilon_F$ at $T=0$ or
$\beta=\infty$ changes as a function of $Q$ or of the interaction
between the particles : $\epsilon_F=\frac{\varepsilon_F}{Q^{2/3}}$
where $\varepsilon_F$ is the conventional Fermi energy at $q=1$ or
$Q=1$. This result can be compared to that of the {\it fractional
exclusion statistics} ($FES$)\cite{Wu94} for intermediate
particles different from bosons and fermions. It is not surprising
to see that nonextensive statistics has similar effect to that of
$FES$ describing low dimension quasi-particles or elementary
excitations\cite{Wu94,Hald91}, because $qed$ is just a result of
interactions which are perhaps the origin of the quasi-particles.
However, the fact that only the $NSM$ formalism with unnormalized
average can give the intermediate quantum distributions seems to
deserve further investigation. Quite interesting efforts have been
made by some authors\cite{Kani96,Raja95,Buyu01} to relate nonextensive
statistics and the quantum distributions given by Eq.(\ref{26}) to
$FES$.

\section{Conclusion}
Summing up, it is argued that the nonextensive thermostatistics
should be based on the factorization of compound probability
suggested, not by ``independence" of noninteracting systems, but
by Abe's pseudoadditivity for $equilibrium$ interacting systems
having Tsallis entropy. So Eq.(\ref{m1}) should be viewed as a
fundamental hypothesis of $NSM$ and has to be rigorously satisfied
by all exact calculations relative to equilibrium systems. All
calculations based on additive hamiltonian Eq.(\ref{1}) are to be
considered as {\it extensive energy approximation} which should be
employed with great care. In this formalism, the applications of
$NSM$ to ideal gases are revisited. The results are different from
those given by extensive approximation. The standard quantum
distributions can be given by using normalized average calculus.
The unnormalized average leads to a kind of fractional exclusion
distributions.

\end{document}